# Surface Transport and Quantum Hall Effect in Ambipolar Black Phosphorus Double Quantum Wells

Son Tran[1,2,†], Jiawei Yang[1,2,†], Nathaniel Gillgren[1], Timothy Espiritu[1], Yanmeng Shi[1], Kenji Watanabe[3], Takashi Taniguchi[3], Seongphill Moon[4,5], Hongwoo Baek[4], Dmitry Smirnov[4], Marc Bockrath[1,2], Ruoyu Chen[1,2\*], Chun Ning Lau[1,2\*]

[1] Department of Physics and Astronomy, University of California, Riverside, CA 92521

[2] Department of Physics, Ohio State University, Columbus, OH 43220

[3] National Institute for Materials Science, 1-1 Namiki Tsukuba Ibaraki 305-0044 Japan.

[4] National High Magnetic Field Laboratory, Tallahassee, FL 32310

[5] Department of Physics, Florida State University, Tallahassee, FL 32310

[†] These authors contribute equally to this work.

**Abstract**

**Quantum wells constitute one of the most important classes of devices in the study of 2D systems. In a double layer QW, the additional "which-layer" degree of freedom gives rise to celebrated phenomena such as Coulomb drag, Hall drag and exciton condensation. Here we demonstrate facile formation of wide QWs in few-layer black phosphorus devices that host double layers of charge carriers. In contrast to tradition QWs, each 2D layer is ambipolar, and can be tuned into n-doped, p-doped or intrinsic regimes. Fully spin-polarized quantum Hall states are observed on each layer, with enhanced Landé g-factor**

[*] Email: chen.7729@osu.edu  
[*] Email: lau.232@osu.edu

**that is attributed to exchange interactions. Our work opens the door for using 2D semiconductors as ambipolar single, double or wide QWs with unusual properties such as high anisotropy.**

Quantum confinement of atoms and electrons to two-dimensions (2D) profoundly alters their electronic properties, giving rise to celebrated phenomena such as the integer and fractional quantum Hall (QH) effect (*1*), quantum spin Hall effects (*2-4*) and interfacial superconductivity and magnetism (*5*), as well as technological development such as QW lasers and photovoltaic solar cells. In double QWs, electrons can reside on either one or both potential wells; this "which-layer" degree of freedom is analogous to spin and often referred to as the pseudospin, thus providing a fascinating platform for investigation of multicomponent physics and competing symmetries(*6-16*).

Traditional QWs are fabricated based on GaAs/AlGaAs heterostructures, and double QWs by inserting a very thin layer of large band-gap material between two adjacent single QWs. More recently, the advent of 2D materials (*17*) has opened the door for manipulating low-dimensional materials systems with an unprecedented level of control. For instance, single QWs based on single- and few-layer graphene (*18, 19*) led to the observation of tunable integer and fractional QH effect (*20-24*), magnetic focusing (*25, 26*), Veselago lensing (*27-31*), and specular Andreev reflection (*32, 33*).

An alternative route to realizing double QWs is to use a wide QW structure (*34-38*), in which, instead of physically inserting a high barrier separating two layers of charges, the Coulomb repulsion among electrons spontaneously causes electrons to reside on the opposite walls, thus forming a double-layer system with a "soft" potential barrier. Compare with double

QWs, wide QWs affords similar physics, except that the soft barrier leads to enhanced interlayer coupling that hybridizes the two layers, giving rise the formation of symmetric and antisymmetric states that are separated by an energy gap $\Delta_{SAS}$. By modulating width of the well $d$, charge density $n$, $\Delta_{SAS}$ and magnetic length $l_B$, inter- and intra-layer Coulomb interactions can be tuned, hence yielding insight into the competing interactions in low dimension systems. Despite the recent progress, such wide QW structures have not been demonstrated in 2D materials to date, due to the lack of high mobility 2D semiconductors with sizeable band gaps.

Black phosphorus, consisting of layered phosphorus atoms in orthorhombic structure (*39, 40*), is such a candidate. A plethora of fascinating properties have been predicted or observed in few-layer BP (*41-45*), such as a layer-, strain- and electric field-dependent band gap (*46-48*), highly anisotropic electrical, thermal and optical properties (*44, 49-52*), and very high mobility (*40, 53-55*) that rival that of first generation of graphene devices. Here we show a simple realization of wide QWs with a double layer of charge carriers in few-layer black phosphorus (BP) devices, with mobility up to 6500 $cm^2$/Vs. Two independent sets of SdH oscillations are observed, indicating the presence of two parallel conducting surface states well separated by an intrinsically insulating bulk. Integer QH plateaus are observed on both surface states. In contrast to traditional GaAs/AlGaAs QWs, BP-based wide QW devices are ambipolar and highly tunable – by independently adjusting top and back gate voltages, we can tune the system into the single or double quantum well regime with unipolar or bipolar charge distributions, which host quantum Hall states at high magnetic fields. From temperature and density-dependent measurements, an enhanced Landé *g*-factor of 2.7 is observed. Our work paves the way for using 2D materials as wide QWs for the investigating phenomena such as Landau level hybridization, inter-well Coulomb interactions, or multi-component QH ferromagnetism (*34, 56, 57*) in this highly anisotropic system.

Here we focus on a device with BP ~ 20 nm thick with Si/SiO$_2$ back gate and hBN/Al$_2$O$_3$/Au top gate at *T*=0.3 K. The schematic of the device is shown in FIG. 1A, and an optical image in FIG. 1B. The devices are measured in a pumped He$^4$ or He$^3$ cryostat. Similar data are observed in multiple devices. FIG. 2A presents the four-terminal resistance *R* of the device (color) as a function of top gate $V_{tg}$ (horizontal axis) and back gate $V_{bg}$ voltages (vertical axis), where green and brown colors indicate conductive (*R*~10$^3$-10$^4$ Ω) and highly resistive (*R*~10$^7$-10$^8$ Ω) states, respectively. Due to the thinner dielectric layers and higher dielectric constant of Al$_2$O$_3$, the coupling efficiency of top gate is approximately 4 times that of the back gate. The white area in the center of the plot corresponds to an insulating regime (*R*>10$^8$ Ω) where the high resistance saturates the amplifier, indicating that the Fermi level is within the band gap. Ambipolar transport can be attained by modulating either $V_{bg}$ or $V_{tg}$: the device is very conductive when highly hole-doped, with high field effect mobility ~6000 cm$^2$/Vs; as either gate is tuned close to 0, its resistance increases precipitously; when the gate voltages are highly positive, electron conduction is turned on, with field effect mobility ~1000 cm$^2$/Vs. Such electron-hole asymmetry in mobility has been observed before (*53, 55, 59*), and attributed to the Cr/Au electrodes that favor contacts to *p*-doped semiconductors. Another notable feature is the the triangular shape of the insulating region: its lower boundary at negative gate voltages (-6.5<$V_{tg}$<3 and -40<$V_{bg}$<-5) moves with a negative slope, indicating that the "on"-state threshold voltage in the hole-doped regime is dependent on the total charge density induced by both gates; in contrast, the boundaries at positive gate voltages ($V_{bg}$ ~14 V and $V_{tg}$~6 V, respectively) remain constant, signifying that the "on"-state threshold voltage in the electron-doped regime is controlled by a single gate. We attribute the flat upper and right boundaries to bulk impurities that give rise to unoccupied localized states within the band gap; as the Fermi level is located very close to the valence band, such states have minimal effect on the transport of BP in the hole-doped regime. As an increasing gate voltage raises the Fermi level towards the conduction band, these unoccupied mid-gap states must first be filled. Thus, back (top) gate cannot effectively switch on the top (bottom) surface state, due to the large number of localized states within the bulk of the BP device. This effect, in combination with Fermi level pinning that reduces the effect of gating, results in the flat boundaries of the insulating region.

Strikingly, at very large positive or negative $V_{tg}$ values, the insulating region almost completely disappears. For instance, at $V_{tg}$=-8 V, the resistance maximum near the charge neutrality point is ~600 kΩ, which is reduced from the global resistance maximum by more than two orders of magnitude. Such a disappearance of the insulating region may arise from the non-trivial closure of the band gap by a large out-of-plane electric field (*47*); alternatively, it could also be a consequence of a non-uniform charge distribution, in which spatially separated surface states contribute to the transport even though the overall net charge is zero.

To determine the origin of the disappearing insulating region, we perform magnetotransport measurements. FIG. 2C plots $R(V_{bg}, V_{tg})$ at magnetic field $B$=18 T, where quantum oscillations are clearly visible. Several different patterns are observed, indicating distinct transport regimes. We first focus on the lower left quadrant: for negative $V_{bg}$ and $V_{tg}$, a checkerboard pattern is observed. The presence of horizontal and vertical sets of oscillations indicates the coexistence of two separate, high mobility 2D hole systems (2DHS), each independently tuned by the adjacent gate. The absence of diagonal features in this quadrant indicates that the effect of the farther gate on each layer is fully screened. Indeed, from a simple

Schrödinger-Poisson calculation, the gate-induced 2D hole wavefunction is expected to tightly confine to the outermost 5 or 6 atomic layers at the surface (*60*), thus limiting the screening length to <4 nm. Thus the device forms a "naked" wide quantum well, with two distinct 2DHS residing at the top and bottom surfaces, separated by an intrinsic or insulating region. Similarly, the upper right quadrant of the figure corresponds to the formation of two distinct 2D electron systems (2DES) that in principle will display similar checkerboard patterns of oscillations at sufficiently high fields. However, due to the relatively low electron mobility, quantum oscillations are not observed at $B$=18T.

In the upper left (lower right) quadrants, *i.e.* when both $V_{tg}$ and $V_{bg}$ have large magnitudes but different signs, only vertical (horizontal) oscillations are observed. Here, similar to the unipolar case, the device hosts top and bottom surface states; however, what distinguishes this case is that the states carry charges of opposite signs. Since the electron-doped regime has lower mobility, quantum oscillations are not observed; thus a single set of oscillations emerges parallel to (i.e. independent of) the axis that corresponds to the farther gate. Hence these quadrants correspond to a wide quantum well with 2DHS and 2DES on opposite surfaces, which has not been realized in GaAs-based devices.

Lastly, when either $V_{bg}$ or $V_{tg}$ are tuned close to 0, we observe only a single set of *diagonal* oscillations, *i.e.* the charge density is controlled by both $V_{tg}$ and $V_{bg}$. In this regime, one of the surfaces is tuned into the intrinsic regime and no longer screens the nearby gate, thus the remaining surface state in this QW is subjected to field lines from both back and top gates. The configurations of the top and bottom surface states that correspond to various regions of the $R(V_{bg}, V_{tg})$ map are summarized in the inset and left panel of FIG. 2C, and the band diagrams in FIG. 2D, where the hole (h)-doped, electron (e)-doped and intrinsic (i) states are represented by red, blue and white regions, respectively. These configurations also establish that the disappearance of the insulating region arises from the formation of 2DHS or 2DES on either surface at large doping.

To sum our experimental observations thus far, we demonstrate that a thin BP device acts as a wide QW, which can host surface states on top and bottom surfaces, whereas the interior is a gapped intrinsic semiconductor that acts as a soft tunnel barrier. Unlike conventional GaAs-based counterparts, these BP-based QWs support both single- and double layer states that are exceedingly tunable, as each of the top and bottom surfaces may be independently tuned to intrinsic, 2D electron gas or 2D hole gas states. These novel wide QWs may be further optimized by improving mobility, or reducing the BP flake thickness and hence the center barrier width so as to enhance Coulomb interactions between the surface states and tuning parameters such as the symmetric-anti-symmetric gap, thus allowing investigation of 2D correlated physics such as 2-component solid, Wigner crystals, interlayer coherence, and reentrant integer and fractional quantum Hall states (*61-63*) with charges of either or both polarities.

Furthermore, we can extract information about the 2D surface states by analyzing the temperature and density dependence of the quantum oscillations. Here we focus on the (p:p) region. FIG. 3A-B plots the background-subtracted resistance $\Delta R(V_{bg})$ at different temperatures and at constant $V_{tg}$=-6 and -4.4V, respectively. The oscillation amplitudes decrease with temperature, and can be fitted within the Lifshitz-Kosevich approach for 2D systems, yielding an effective mass $m^*$~0.43 ± 0.1 $m_e$, where $m_e$ is the electron rest mass. This value is in good agreement with that obtained from density functional theory (DFT) calculations (*53*). Notably, we find no clear density dependence of $m^*$ within error bars.

A close examination of FIG. 3A-B reveals a salient feature: the oscillation amplitude is not monotonic in density; at some gate voltages the peak height alternates between adjacent oscillations at some gate voltages, as indicated by the arrows. Such non-monotonic and/or alternating peak heights are not expected in conventional quantum oscillations, where the equally spaced Landau levels at constant $B$ yield oscillation amplitudes that scale as $n^{-1/2}$. They have been observed in a number of systems, such as ZnO heterostructures (*64*), Si inversion layers (*65, 66*), SrTiO$_3$ (*67*) and more recently, in thin BP sheets (*53, 55*), and are commonly attributed to the appearance of the Zeeman gap that is smaller than the single particle cyclotron gaps, where the oscillation amplitude is given by (*67*)

$$\frac{\Delta R}{R_0} = \frac{5}{2} \sum_{s=1}^{\infty} b_s \cos\left(\frac{2\pi n h}{2eB} s - \frac{\pi}{4}\right) \qquad (1)$$

$$b_s = \frac{(-1)^s}{\sqrt{s}} \left(\frac{eB}{nh}\right)^{1/2} \frac{2\pi^2 s k_B T/\hbar\omega_c}{\sinh(2\pi^2 s k_B T/\hbar\omega_c)} \exp\left(\frac{-2\pi^2 s k_B T_D}{\hbar\omega_c}\right) \cos\left(\frac{\pi s g m^*}{2 m_e}\right)$$

where $\hbar$ is the reduced Planck constant, $g$ the Lande g-factor, $n$ the carrier density, $\omega_c = eB/m^*$ the cyclotron frequency, $m^*$ is the effective mass, $k_B$ the Boltzmann constant, $T_D$ the Dingle temperature, and $s = 1, 2$. In these equations, the periodicity of the oscillations is controlled by $n$, and our data yield a capacitive coupling ~6.5 x $10^{10}$ cm$^{-2}$ V$^{-1}$ between the back gate and the bottom surface state. The ratio between the alternative peak heights is controlled by the combined product $gm^*$. By fitting $\Delta R(V_{bg})$ curves to Eq. (1), we obtain good agreement by using $T_D$=2 K and $gm^*$=1.15±0.05 (FIG. 3D). As we have determined $m^*$~0.43 from the temperature dependence of the oscillations, the fitting results indicate the Landé g-factor is ~ 2.7, which represents a ~33% enhancement over the free hole value of 2.0. Such enhancement likely originates from the exchange interaction among electrons in spin-polarized Landau levels (*68-70*).

Finally, at sufficiently high magnetic field, the quantum Hall effect, which is a prototypical 2D phenomenon, can be observed on both 2DHS. FIG. 4A exhibits the Landau fan $R(V_{bg}, B)$ at $V_{tg}$=0 V for 18<$B$<31T, and several line traces are shown in FIG. 4B. At $V_{tg}$=0 V, the top layer is turned off and only the bottom layer participates in electrical transport. Quantized plateaus at filling factor $v$=1, 2, 3, 4 and 5 are observed, indicating full lifting of the spin degeneracy. On the other hand, at $V_{tg}$=-8V, the $R(V_{bg},B)$ data exhibit additional vertical strips superimposed on top of the Landau fan, signifying the presence of quantum Hall states on the highly hole-doped top layer (FIG. 4C), with an estimated hole density of $2.4 \times 10^{12}$ cm$^{-2}$. No quantized plateau is observed in the raw data (FIG. 4D, dashed lines), due to the co-existence of QH states on both top and bottom surfaces. As $B$ sweeps from 18 to 31T, the filling factor of the top surface state is estimated to decrease from $v$=-4.5 to -2, thus we model its conductance as stepwise quantized plateaus at appropriate filling factors (FIG. 4C inset). By subtracting this calculated parallel conductance from the raw data, plateaus are recovered in $R(V_{bg})$ data, similar to those in FIG. 4A-B. Take together, these results indicate that both top and bottom 2DHS host quantum Hall states, with spin degeneracy fully lifted.

In conclusion, we have demonstrated the formation of ambipolar, highly tunable wide QWs in hBN-encapsulated dual-gated BP devices, in which either or both surfaces may be tuned into intrinsic insulator, or hole-doped or electron-doped surface states. At high magnetic fields, fully spin-resolved integer quantum Hall states are observed in both top and bottom surface states. By further optimization such as mobility improvement and thickness reduction, these BP-based wide QW devices may open the door to a wide range of novel physics in this highly anisotropic 2D systems, ranging from Wigner crystallization and interlayer coherence to integer and fractional quantum Hall states with possible reentrant behavior.

**Materials and Methods**

Bulk hBN and BP crystals are grown via high temperature and high pressure techniques (*39*), and exfoliated into thin sheets onto Si/SiO$_2$ substrates. A dry transfer technique is used to assemble hBN/BP/hBN stacks (*58*) inside a VTI glove box with moisture and oxygen concentration <0.1 ppm. The top hBN layer is etched in SF$_6$ plasma to expose the BP layer, and Cr/Au electrodes are deposited thereafter by electron beam evaporation. To fabricate the top gate, a dielectric layer of 50-70 nm Al$_2$O$_3$ is deposited onto the entire stack. The devices are measured in a He$^3$ cryostat using standard dc or lock-in techniques in National High Magnetic Field Lab at magnetic fields ranging from 0 T to 30 T.

**Acknowledgement**

This work is supported by FAME center, one of six centers of STARnet, a Semiconductor Research Corporation program sponsored by MARCO and DARPA, and by NSF/ECCS 1509958. A portion of this work was performed at the National High Magnetic Field Laboratory, which is supported by National Science Foundation Cooperative Agreement No. DMR-1157490 and the State of Florida. K.W. and T.T. acknowledge support from the Elemental

Strategy Initiative conducted by the MEXT, Japan and JSPS KAKENHI Grant Numbers JP26248061, JP15K21722 and JP25106006.

**Competing Interests**

The authors declare that they have no competing interests.

**Data Availability**

All data needed to evaluate the conclusions in the paper are present in the paper and/or the Supplementary Materials. Additional data related to this paper may be requested from the authors.

**Author Contribution**

RC and CNL designed the experiment. TT and KW synthesized BP and hBN crystals. JY, ST, TE, YS and NG fabricated samples. JY, ST, NG, SM, HB and DS performed transport measurements. JY, ST, RC, MB and CNL analyzed and interpreted data. JY, ST, RC and CNL wrote the manuscript. All authors discussed and commented on the manuscript.

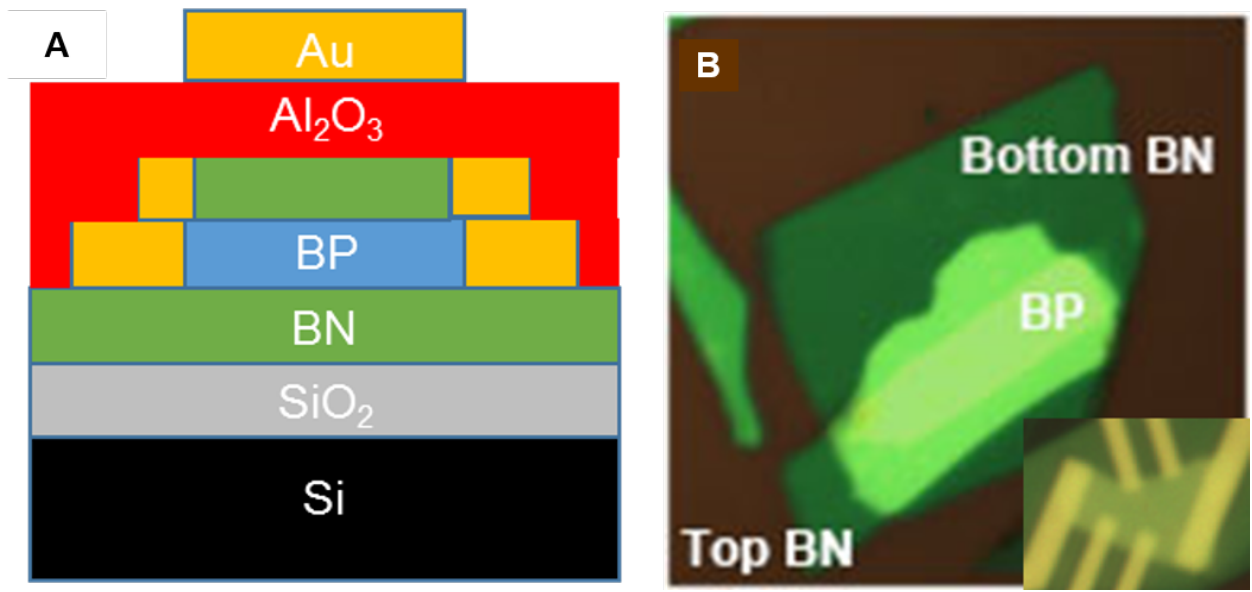

**FIG. 1. Device schematics and image. (A).** Side view of device schematics. **(B).** Optical microscope image of an hBN/BP/hBN stack and of a finished device without top gate (inset).

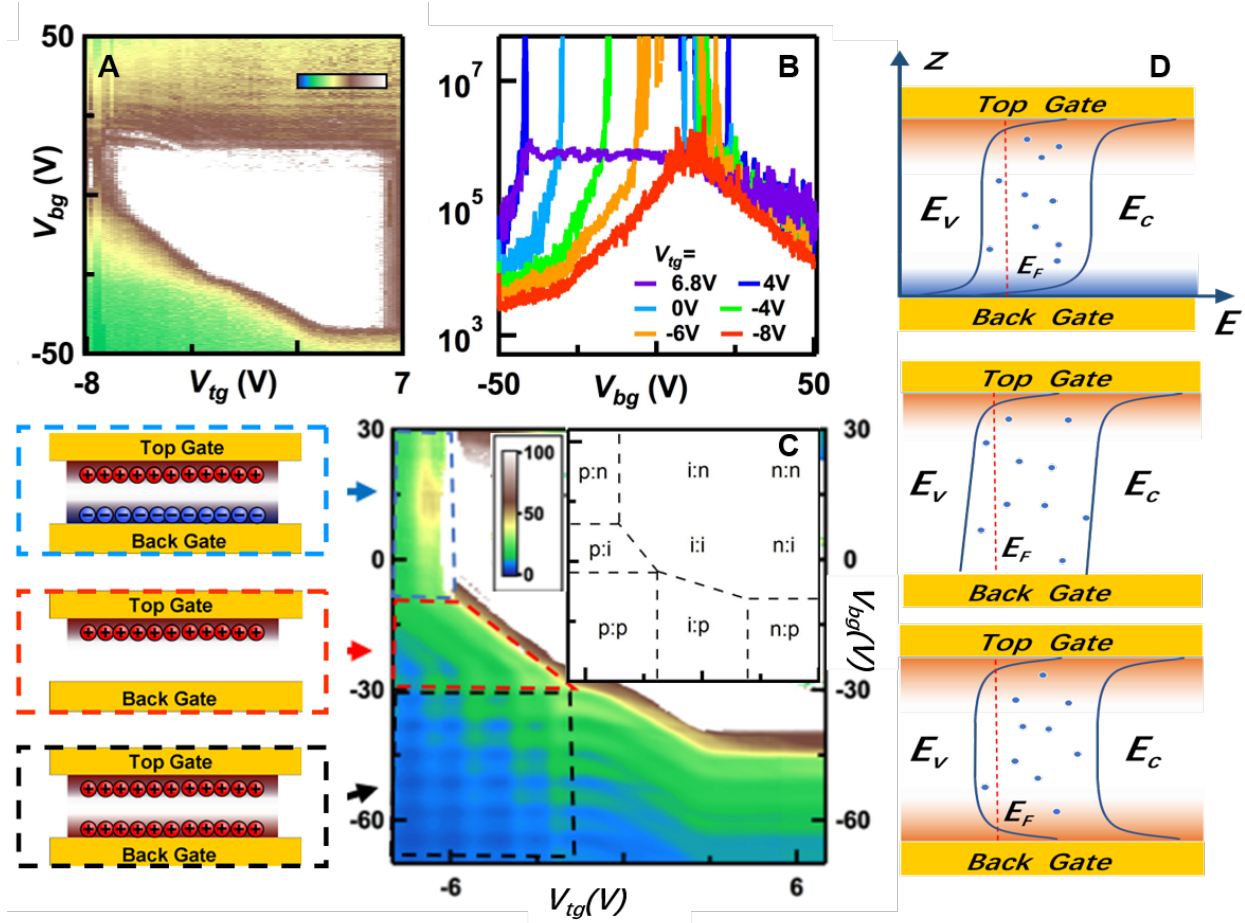

**FIG. 2. Transport data. (A-B).** $R(V_{bg}, V_{tg})$ and line traces $R(V_{bg})$ at different $V_{tg}$ at $T$=1.7 K and $B$=0. Note the logarithmic color scale (in Ω). **(C).** Right panel: $R(V_{bg}, V_{tg})$ at $T$=0.5 K and $B$=18 T, featuring a complicated quantum oscillations pattern. The color scale is in kΩ. Left panels: schematics of the charge distributions that correspond to bipolar double layer, single layer, and unipolar double-layer regimes, respectively. (inset). Charge types for top and bottom surfaces at different combinations of gate voltages. p: hole doped; n: electron doped i: intrinsic insulating state. **(D).** Band diagrams that correspond to the three regimes in (C), with dots illustrating mid-gap impurity states.

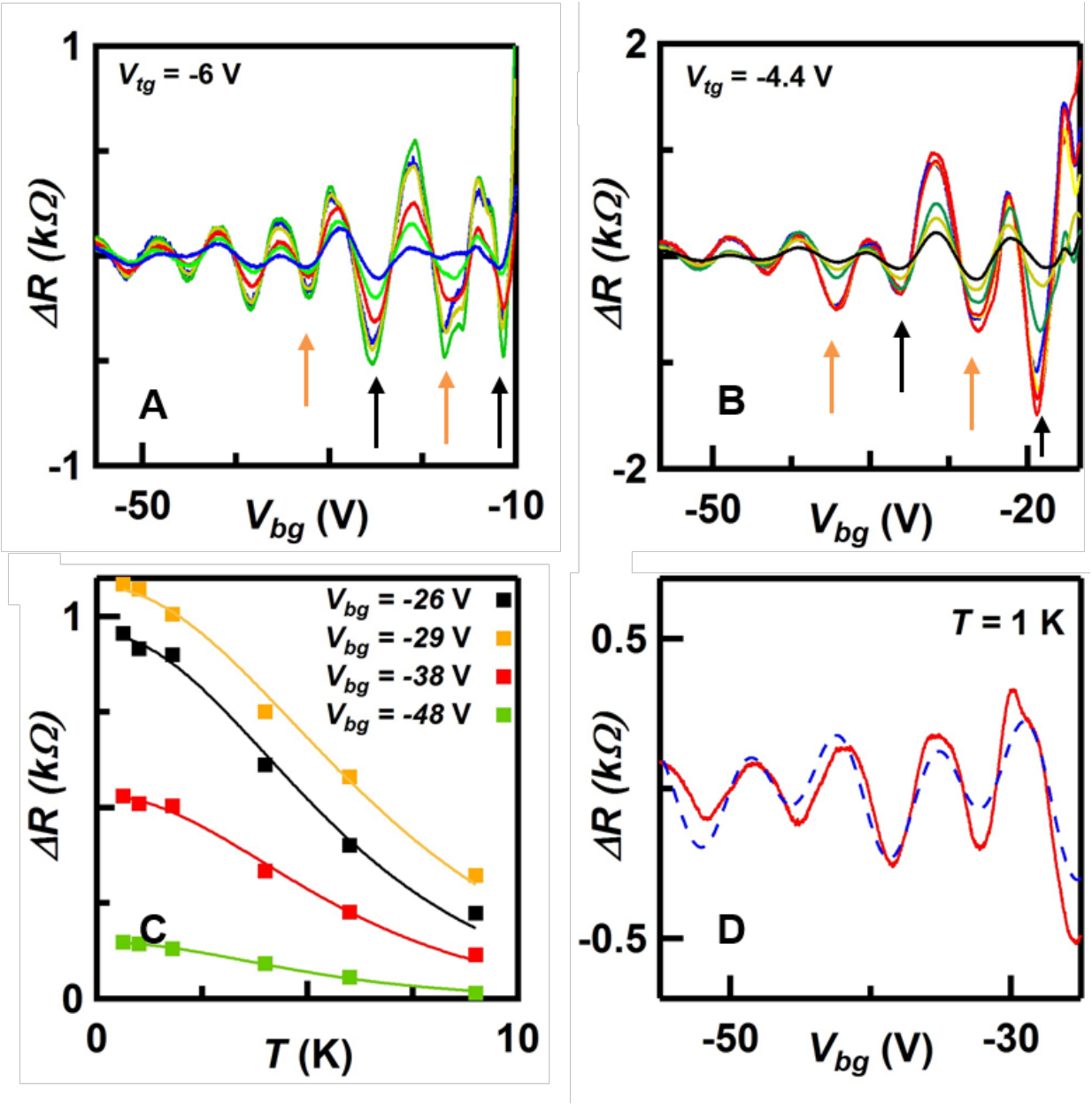

FIG. 3. Quantum oscillations at *B*=18T. (A). Background subtracted resistance $\Delta R$ at $V_{tg}$=-6V and *T*= 0.6, 1, 1.8, 4, 6, and 9K, respectively. Arrows indicate non-monotonic amplitude dependence on density. (B). Similar data set at $V_{tg}$=-4.4V. (C). Oscillation amplitude as a function of temperature at $V_{tg}$=-3V and different $V_{bg}$ values (squares), fitted to Lifshitz-Kosevich formula (solid lines). The fits yield an effective mass $m^* \sim 0.43 \pm 0.1$ $m_e$. (D). $\Delta R(V_{bg})$ at $V_{tg}$=0 and *T*=1K (solid lines), and fitted curve using Eq. (1) (dashed line), $T_D$=2 K and $gm^*$=1.15.

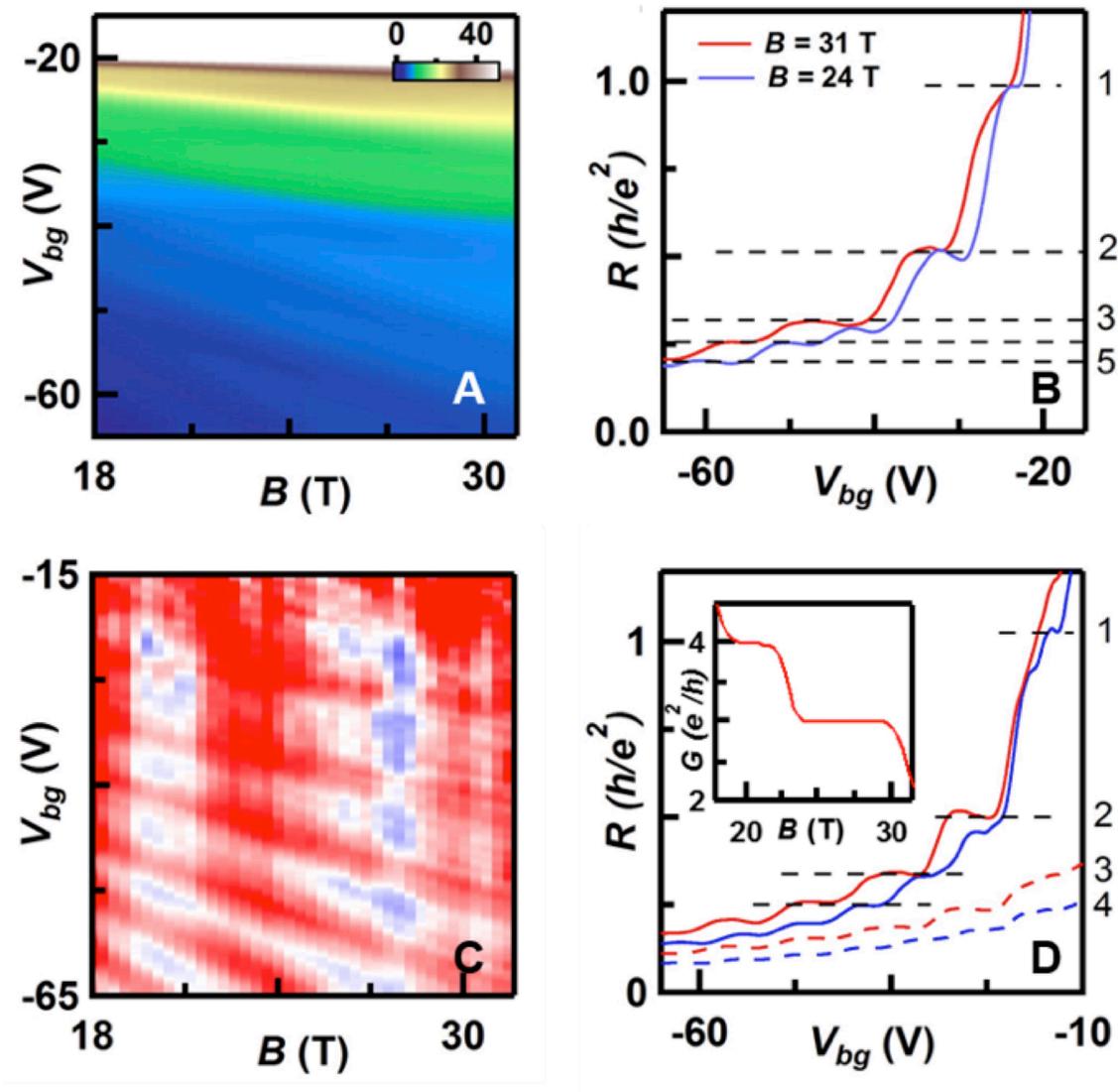

**FIG. 4. Quantum Hall states at high magnetic field. (A).** $R(V_{bg}, B)$ with top gate disconnected for 18<B<31T. Quantized plateaus at filling factors $\nu$ = 1, 2, 3, 4, 5 and 6 are observed. **(B).** Line traces of panel **(A)** at $B$=24T (blue) and 31T (red), respectively. The dashed lines mark expected values of resistance plateaus. **(C).** Differentiated $dR/dB(V_{bg}, B)$ at $V_{tg}$=-8V. **(D).** Line traces of panel **(D)** at $B$=20.3T (blue) and 27.7T (red). The dashed lines correspond to raw data, and solid

lines are obtained by subtracting parallel conductance contributed by top surface states. Inset: calculated conductance of the top surface states at $V_{tg}$ as a function of $B$.